\documentclass{article}
\usepackage{spconf,amsmath,graphicx,hyperref}
\usepackage{booktabs,multirow}

\usepackage{makecell}
\usepackage{caption}
\usepackage{subcaption}
\title{Acoustic Honeybee Queen-State Detection Under Unseen Conditions}
\name{Mahsa Abdollahi$^{1}$, Nico Coallier$^{2}$, Maxime Fraser Franco$^{2}$, Tiago H. Falk$^{1}$}
\address{
$^{1}$INRS-EMT, Université du Québec, Montréal, Canada\\
$^{2}$Nectar Technologies, Montréal, Canada
}

\begin{document}
%
\maketitle
\begin{abstract}
Honeybee queen loss is a major threat to colony health, yet queen-status assessment remains largely manual and disruptive. Acoustic monitoring offers a non-invasive alternative by enabling continuous analysis of hive sounds. In this paper, we benchmark conventional and learned acoustic representations for automated detection of queen absence, comparing task-specific convolutional neural networks with pretrained audio transformers. Experiments are performed on 5,129 audio recordings from 3,285 hives across 47 apiaries collected between October 2024 and August 2026, evaluated using hive- and apiary-independent splits. Results show that models relying on modulation spectrograms achieve the best performance, reaching an Area Under the Receiver Operating Characteristic (AUROC) of 0.81 and Area Under the Precision-Recall Curve (AUPRC) of 0.37 on unseen hives; performance decreases under unseen-apiary evaluation. Overall, our results highlight the promise of modulation-based audio representations for non-invasive queen-status monitoring and highlight the challenge in cross-apiary model generalization.

\end{abstract}
\section{Introduction}
\label{sec:intro}
Domestic honeybees are among the most important pollinators in global agriculture, underpinning the yield of roughly three-quarters of the world's leading food crops and an ecosystem service valued at USD 235--577 billion annually. As such, colony losses threaten both food security and beekeeping livelihoods~\cite{potts2016safeguarding}. Within a colony, productivity depends strongly on the queen, whose loss disrupts brood production and pheromone signaling, ultimately leading to colony decline if undetected. Despite its importance, queen-status assessment today still relies heavily on time-consuming manual and disruptive hive inspections~\cite{ruvinga2023identifying}, motivating the need for continuous, non-invasive monitoring. Acoustic monitoring is particularly promising because hive sounds reflect collective colony activity and can be captured continuously with low-cost microphones. Early studies showed that queenright and queenless colonies exhibit distinguishable acoustic signatures~\cite{nolasco2019audio,soares2022mfcc,ruvinga2023identifying}. 

Initial approaches paired hand-crafted features, such as mel-frequency cepstral coefficients (MFCCs) and spectral descriptors with classical classifiers~\cite{soares2022mfcc,ho2023evaluating} to predict queen state. More recent work has introduced the use of deep neural networks, including convolutional neural networks (CNNs), recurrent and multilayer neural models~\cite{ruvinga2023identifying}. Despite these advances in the machine learning side, MFCCs and mel-spectrograms have remained effective as inputs for short-duration queen-status classification under noisy field conditions~\cite{ormeno2026acoustic}.

Beyond conventional time-frequency representations, modulation spectral analysis has emerged as a complementary approach for remote beehive monitoring, capturing temporal dynamics not explicitly available in conventional spectrograms. Modulation spectral features and representations have been used recently for colony population prediction and Varroa (\textit{Varroa destructor}) mite infestation detection~\cite{abdollahi2025audio,abdollahi2026prediction}, with improved cross-hive generalization capability~\cite{abdollahi2026improved}. In parallel, self-supervised learning (SSL) has enabled audio encoders pretrained on large-scale corpora~(e.g., \cite{abdollahi2025urban,zhu2024mspb}) to learn general-purpose representations that may capture acoustic characteristics difficult to obtain from conventional descriptors. Guimarães \textit{et al.} recently benchmarked self-supervised and supervised audio representations across multiple beehive-monitoring tasks, demonstrating their potential for extracting informative features from hive sounds~\cite{guimaraes2025benchmarking}.

Despite this progress, two important gaps remain. First, generalization across unseen hives, apiaries (yards), and recording environments is still unresolved. Existing queen-status studies often involve few colonies, and performance can drop substantially on previously unseen hives~\cite{nolasco2019audio,sad2025deep}. Recent work, however, using short segments and multiple acoustic representations has reported performance within its own dataset splits rather than isolating hive-level generalization~\cite{ormeno2026acoustic}. Second, the effectiveness of pretrained audio representations for queenright/queenless classification remains largely unexplored, particularly under unseen apiary conditions, leaving it unclear whether they offer any advantage over conventional acoustic features and e.g., task-specific CNNs. In this work, we address these gaps by systematically evaluating conventional acoustic representations, CNN-based models, and pretrained audio representations for acoustic queen-status classification. We place emphasis on hive-independent evaluation to determine whether learned representations generalize across colonies rather than capturing colony-specific acoustic signatures.

\section{PROPOSED APPROACH}

We investigate whether queen state can be inferred from acoustic hive recordings and compare the use of different acoustic representations and classifiers. The sections to follow provide more details on these two methods, as well as on the fusion of contextual metadata, including recording time and location, projected directly into the learned acoustic embedding.

\subsection{Acoustic representations}
\label{ssec}

We evaluated four acoustic representations: MFCCs (with delta and double-delta MFCCs), mel spectrograms, multi-resolution mel spectrograms, and modulation spectrograms. All recordings were sampled at 16 kHz and analyzed using Hann windows. The input shapes are summarized in Table~\ref{tab:features}.

\noindent\textbf{MFCCs:} We extracted MFCCs using a 64-band mel filterbank and appended their first- and second-order temporal derivatives (delta and double-delta, respectively), providing a representation that captures both spectral characteristics and their temporal evolution.

\noindent\textbf{Mel spectrograms:} Log-mel spectrograms were computed using a relatively long analysis window to provide increased frequency resolution. Frequencies up to 4 kHz were retained to capture the bee wingbeat fundamental and its harmonics.

\noindent\textbf{Multi-resolution mel spectrograms:} 
Log-mel spectrograms were computed with three analysis windows (1024, 2048 and 4096
samples, i.e.\ 64, 128 and 256\,ms) sharing the same hop and mel filterbank, and stacked
as input channels. Shorter windows smear transients less in time, whereas longer windows
resolve narrow spectral peaks more sharply; stacking lets each convolution combine both
at every time--frequency location. 

\noindent\textbf{Modulation spectrograms:} We employed a two-stage spectral analysis to characterize temporal fluctuations within individual mel bands. First, a mel magnitude spectrogram was computed, producing one temporal envelope per frequency band. After mean removal, each envelope was analysed in the modulation-frequency domain over short temporal segments. Low-frequency modulation components were retained and pooled into logarithmically spaced modulation bands.
We considered two modulation representations. The \emph{averaged} representation aggregates modulation information over time to produce a compact frequency--modulation map. The \emph{framed} representation preserves the temporal sequence of modulation windows and is used as the default configuration unless stated otherwise. The interested reader is referred to \cite{abdollahi2025audio, abdollahi2026improved, abdollahi2026prediction} for more details on the computation of the modulation spectral signal representation.

\begin{figure*}
\centering
\subfloat[]{\label{fig:}
\centering
\includegraphics[width=0.25\linewidth]{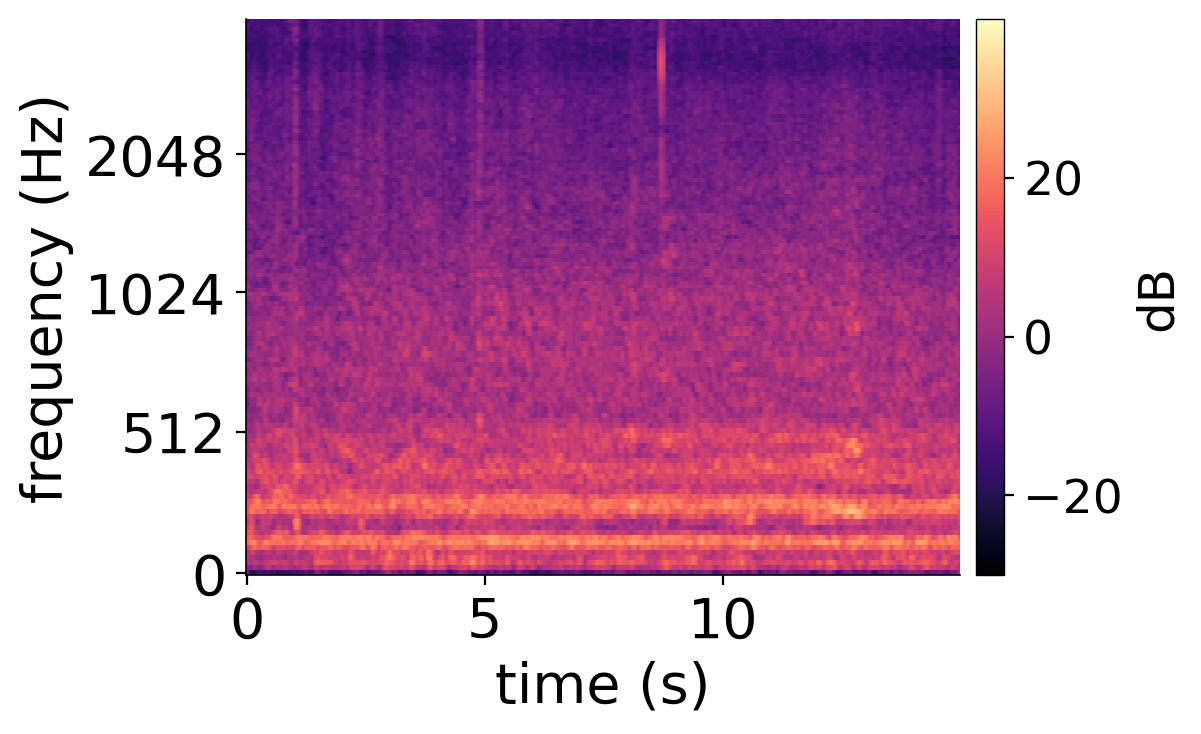}
}
\subfloat[]{\label{fig:}
\centering
\includegraphics[width=0.25\linewidth]{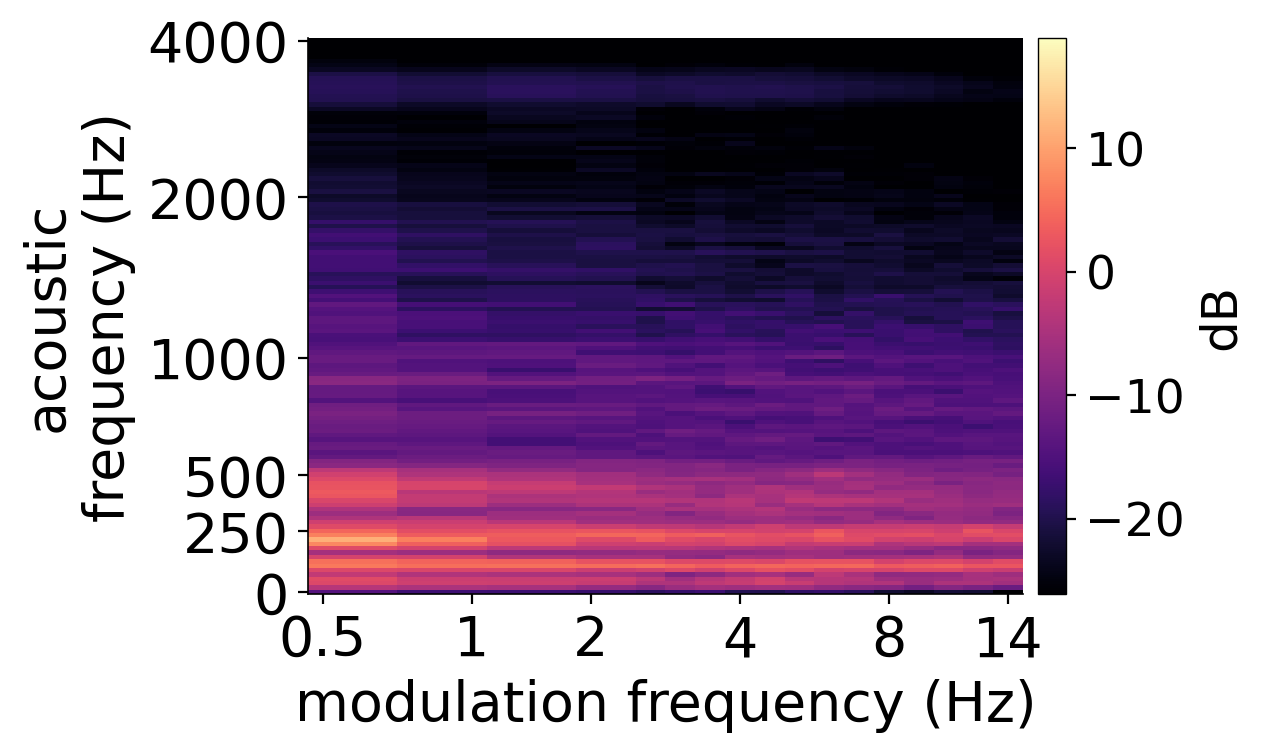}
}
\subfloat[]{\label{fig:}
\centering
\includegraphics[width=0.25\linewidth]{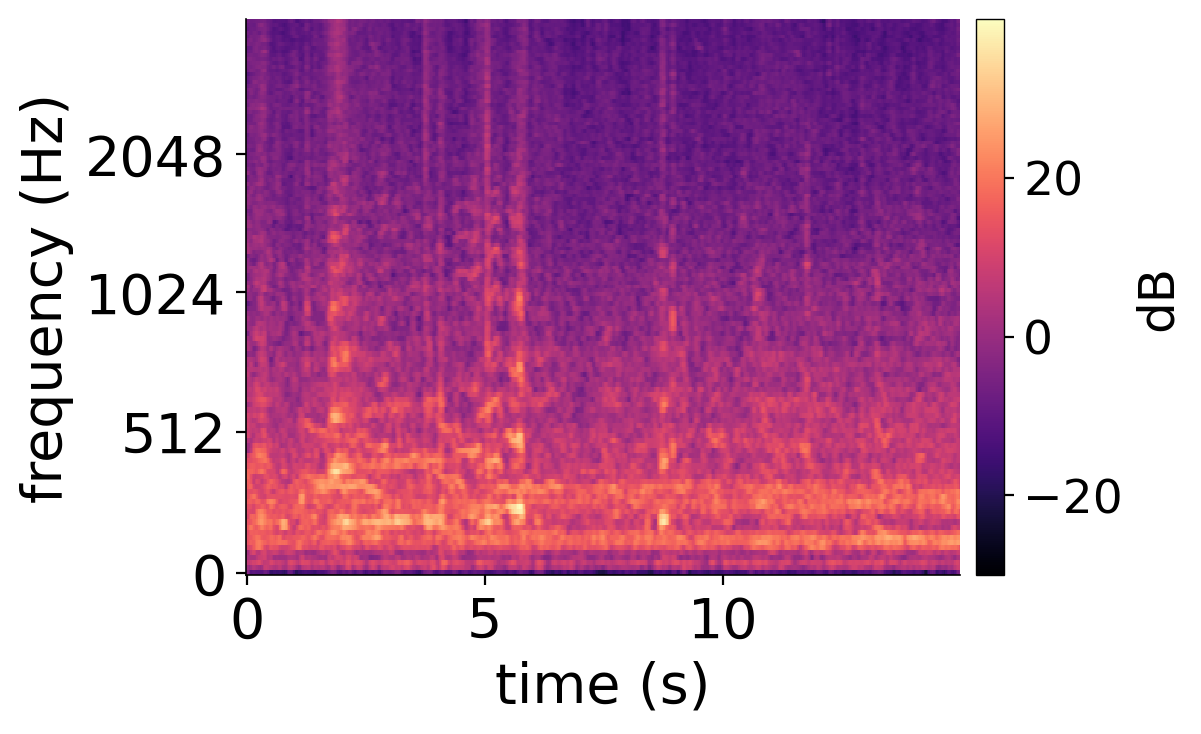}
}
\subfloat[]{\label{fig:}
\centering
\includegraphics[width=0.25\linewidth]{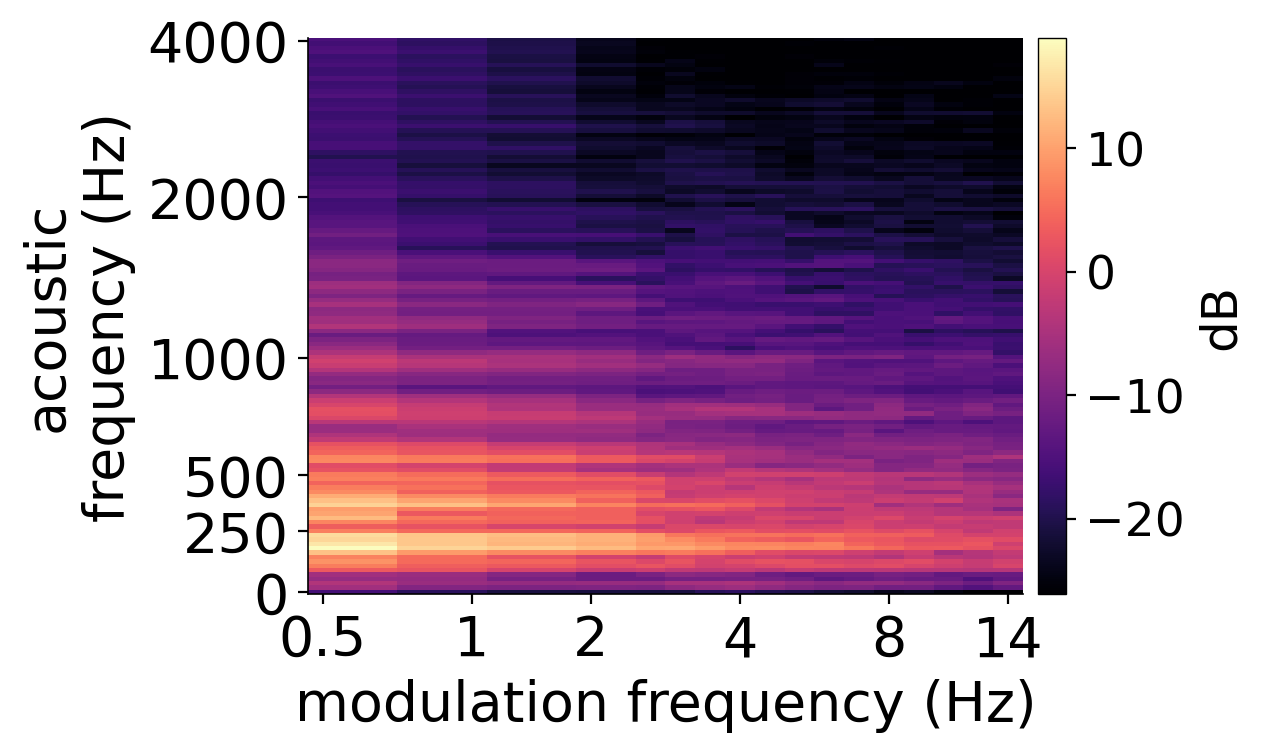}
}
\caption{Matched queenright/queenless recordings from two hives in the same operation: (a) queenright mel-spectrogram, (b) queenright modulation spectrogram, (c) queenless mel-spectrogram, and (d) queenless modulation spectrogram.}
\label{fig:mel_mod}
\end{figure*}

\begin{figure}
\centering
\subfloat[]{\label{fig:}
\centering
\includegraphics[width=0.5\linewidth]{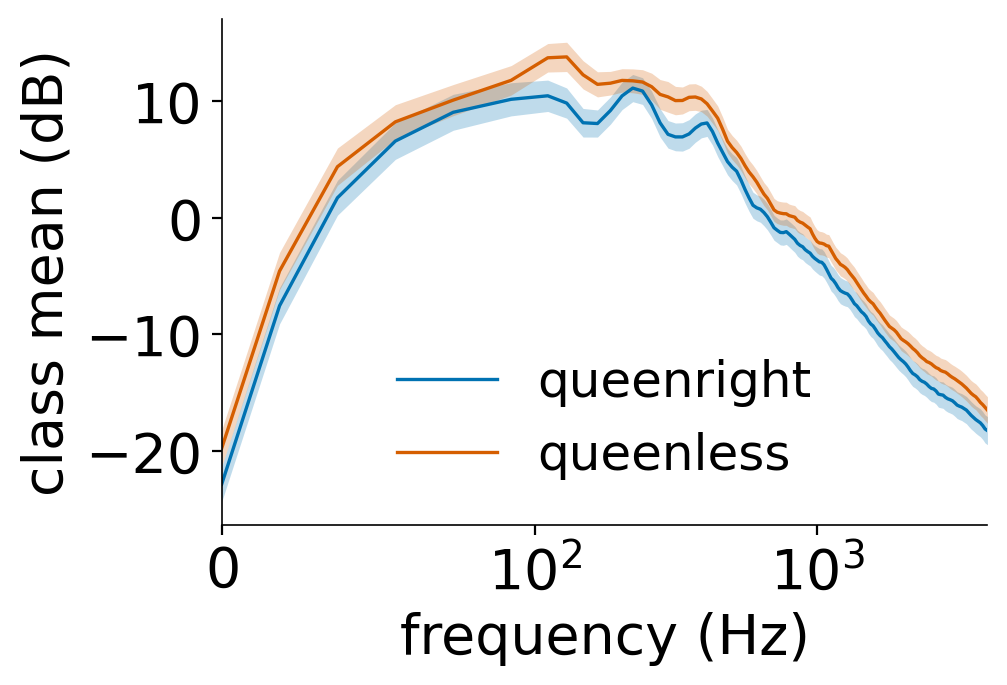}
}
\subfloat[]{\label{fig:}
\centering
\includegraphics[width=0.5\linewidth]{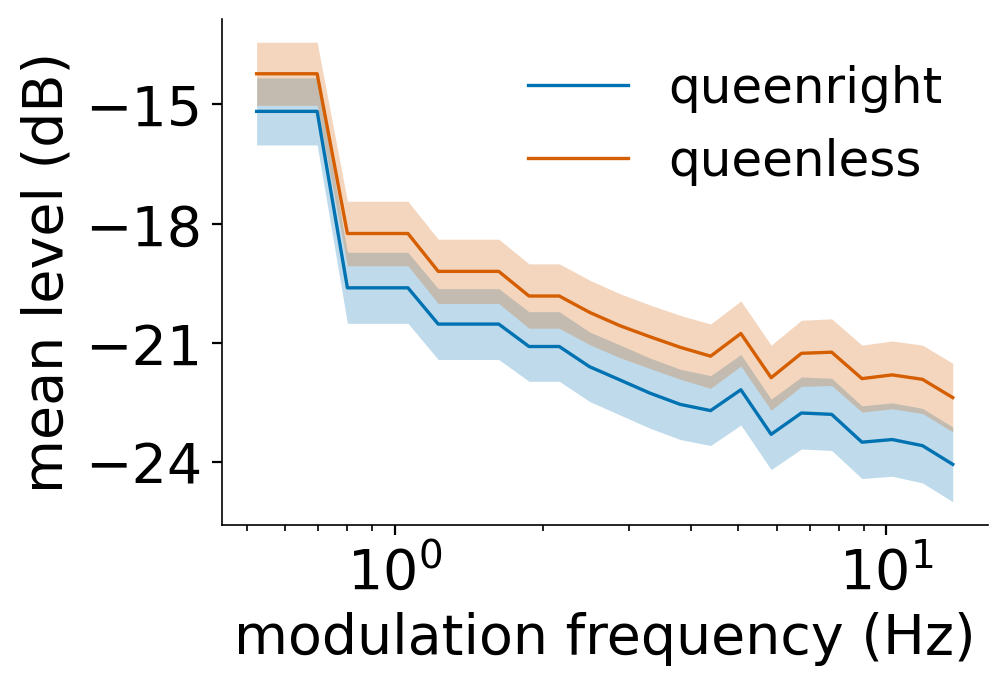}
}
\caption{Class-average profiles of clip pairs
matched on date, time of day and colony size. (a)~Time-averaged log-mel spectrum.
(b)~Clip-mean modulation profile. Shading: 95\% CI of the class mean.}\label{fig:avg_profiles}
\end{figure}

Figure~\ref{fig:mel_mod} compares a queenless and a queenright colony of similar size, recorded minutes apart in the same apiary. The queenright colony produces
a steady, narrow-band hum below 500~Hz. The queenless colony is louder and less stationary, with broadband transients up to 1--2~kHz and more slow modulation
($<2$~Hz) across 200--1000~Hz. Averaged over 160 matched pairs (see Fig.~\ref{fig:avg_profiles}), queenless colonies are louder in every mel band, by about 3~dB below 150~Hz and 1~dB at the wingbeat peak. 

\begin{table}
\centering
\caption{Acoustic representations. All share a 512-sample hop,
$f_{\max}=4$\,kHz and $T=469$ frames per 15\,s clip.}
\label{tab:features}   
\small
\setlength{\tabcolsep}{3pt}
\begin{tabular}{@{}lcl@{}}
\toprule
Representation & $n_{\mathrm{FFT}}$ & Input shape \\
\midrule
MFCC $+\,\Delta,\Delta^2$ & 2048 & $192\times T$ \\
Mel spectrogram           & 4096 & $128\times T$ \\
Multi-res mel               & 1024/2048/4096 & $128\times T\times3$ \\
Modulation (avg.)         & 1024 & $128\times24$ \\
Modulation (framed)       & 1024 & $128\times13\times24$ \\
\bottomrule
\end{tabular}
\end{table}

\subsection{Classification}
\label{ssec:cnn}
All acoustic representations were classified using the same CNN architecture to enable a consistent comparison across input features. The network is based on a truncated YAMNet-style MobileNetV1 backbone~\cite{hershey2017cnn, howard2017mobilenets}, composed of depthwise-separable convolutional blocks and producing a compact acoustic embedding. Two-dimensional representations were provided with a singleton channel dimension, whereas multi-resolution representations were treated as multi-channel inputs.
Inputs were standardized using statistics computed from the training split. During training, Gaussian noise and small temporal shifts were applied as data augmentation to improve robustness to recording variability and temporal misalignment.
The final convolutional feature map is aggregated with generalized mean (GeM) pooling~\cite{radenovic2018fine}, with a single exponent $p$ learned jointly with the network.

To complement the CNN comparison, we fine-tuned two pretrained audio transformers, each with its own log-mel front-end and therefore operating
directly on raw waveforms rather than the representations in Section~\ref{ssec}. Both transformers use the same class-balanced loss, split protocol, and
model-selection criterion as the CNN.

\noindent\textbf{AST:} The Audio Spectrogram Transformer (AST)~\cite{gong2021ast}
was pretrained on AudioSet in a supervised manner. We fine-tune the full encoder
and represent each clip by the average of its two output summary tokens
([CLS] and distillation), a 768-d vector, replacing the 527-class head with a
randomly initialized two-way linear layer.

\noindent\textbf{MSM-MAE:} The masked-spectrogram masked autoencoder
(MSM-MAE)~\cite{niizumi2022masked} was pretrained in a self-supervised manner without
labels. Since its encoder window is shorter than a clip, frame-level features
are mean-pooled into a single clip embedding, followed by a small multi-layer perceptron.

\noindent\textbf{Metadata fusion:} Each recording is associated with time and geographic location, encoded as periodic sinusoidal and continuous positional features, respectively. These are projected into compact embeddings and concatenated with the acoustic embedding before classification. Fusion occurs after pooling, leaving the acoustic encoders unchanged across representations and backbones.

\section{Experimental Setup}
\subsection{Dataset}

The dataset is drawn from the proprietary audio archive of a commercial hive-management
platform, Nectar Technologies Inc, Canada (\url{https://www.nectar.buzz/})), where beekeepers record short colony clips during routine inspections
and log the inspection outcome in the same record. Recordings are therefore
collected in-the-wild using hand-held devices and may contain wind, traffic,
speech, and other environmental noise. Each recording is a $15$\,s mono clip
resampled to $16$\,kHz.

The cleaned dataset contains 5,129 recordings with known queen state, collected
from 3,285 hives across 47 beekeeping apiaries, most of which are located in
North America, between October 2024 and
August 2026; 10.9\% are queenless. The binary target is
\emph{queenright} ($y{=}1$; laying queen present) or \emph{queenless}
($y{=}0$). Labels are based on the beekeeper-entered hive status.
When an automated
audio grade is available, the beekeeper's entry takes precedence; an automated
label is retained only when it was reviewed and not contested. Recordings
without a valid queen-state label are assigned $-1$ and excluded.

\subsection{Testing Setup and Figures-of-Merit}

We evaluate generalization under two group-independent settings. In the \textit{hive-independent} setting, recordings from each hive are assigned exclusively to one split. In the \textit{apriary-independent} setting, entire operations and their hives are assigned to a single split, leaving validation and test operations unseen during training. These splits limit leakage from shared recording conditions and environments.
Table~\ref{tab:cnn} summarizes all system configurations. CNN variants share the same training setup and differ only in the input tensor (Table~\ref{tab:features}), while the pretrained transformers use their respective front-ends and optimizer settings. Class imbalance ($10.9\%$ queenless) is addressed using balanced cross-entropy weights, and the checkpoint evaluated on test is always the best-validation-loss epoch.
We report Area Under the Receiver Operating Characteristic (AUROC), queenless-focused Area Under the Precision-Recall Curve (AUPRC), precision and recall ($\text{Prec.}_{\text{ql}}$, $\text{Rec.}_{\text{ql}}$) and queenright recall ($\text{Rec.}_{\text{qr}}$) to assess discrimination and performance on both classes.

\begin{figure}
\centering
\subfloat[]{\label{fig:}
\centering
\includegraphics[width=0.5\linewidth]{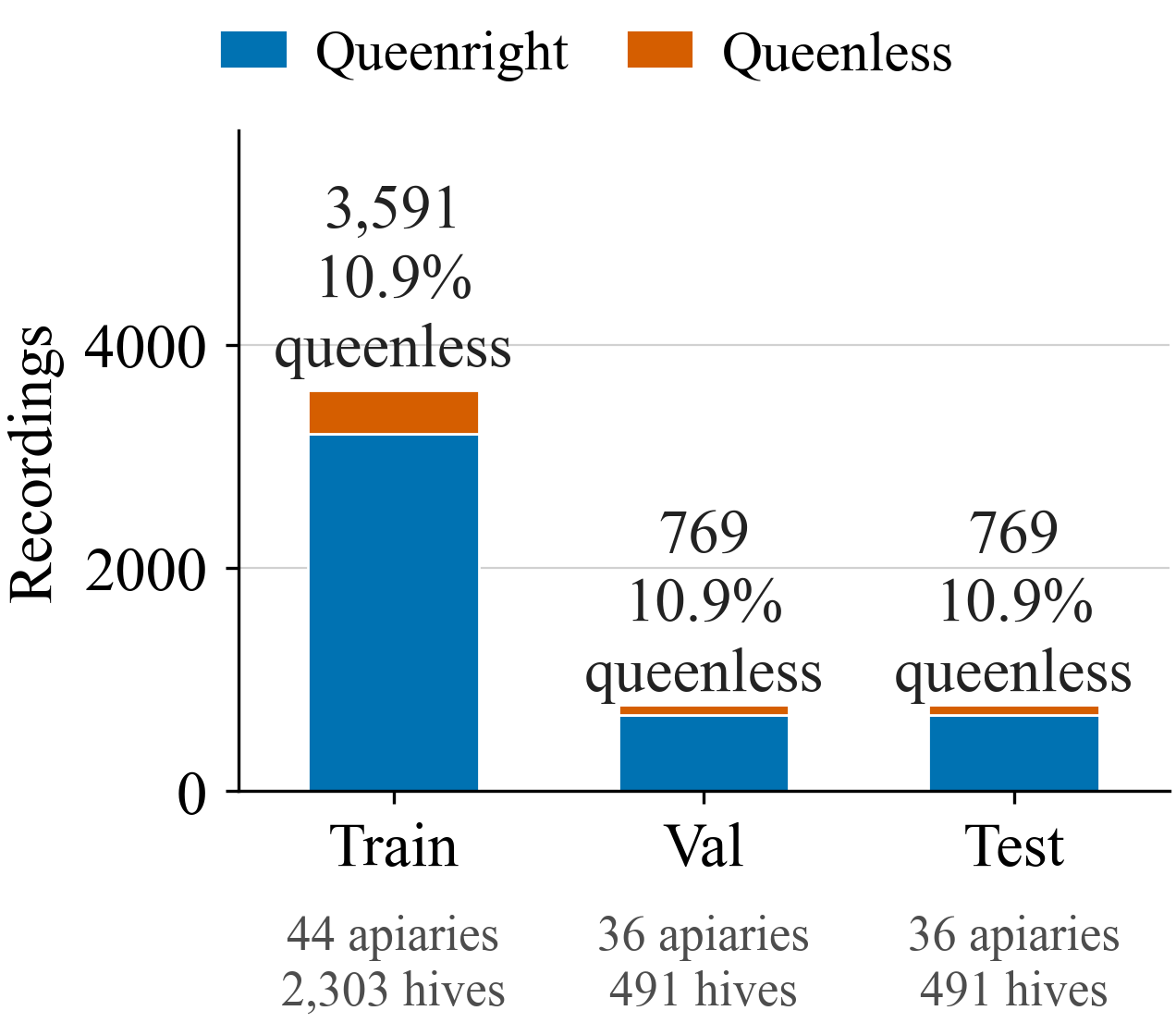}
}
\subfloat[]{\label{fig:}
\centering
\includegraphics[width=0.5\linewidth]{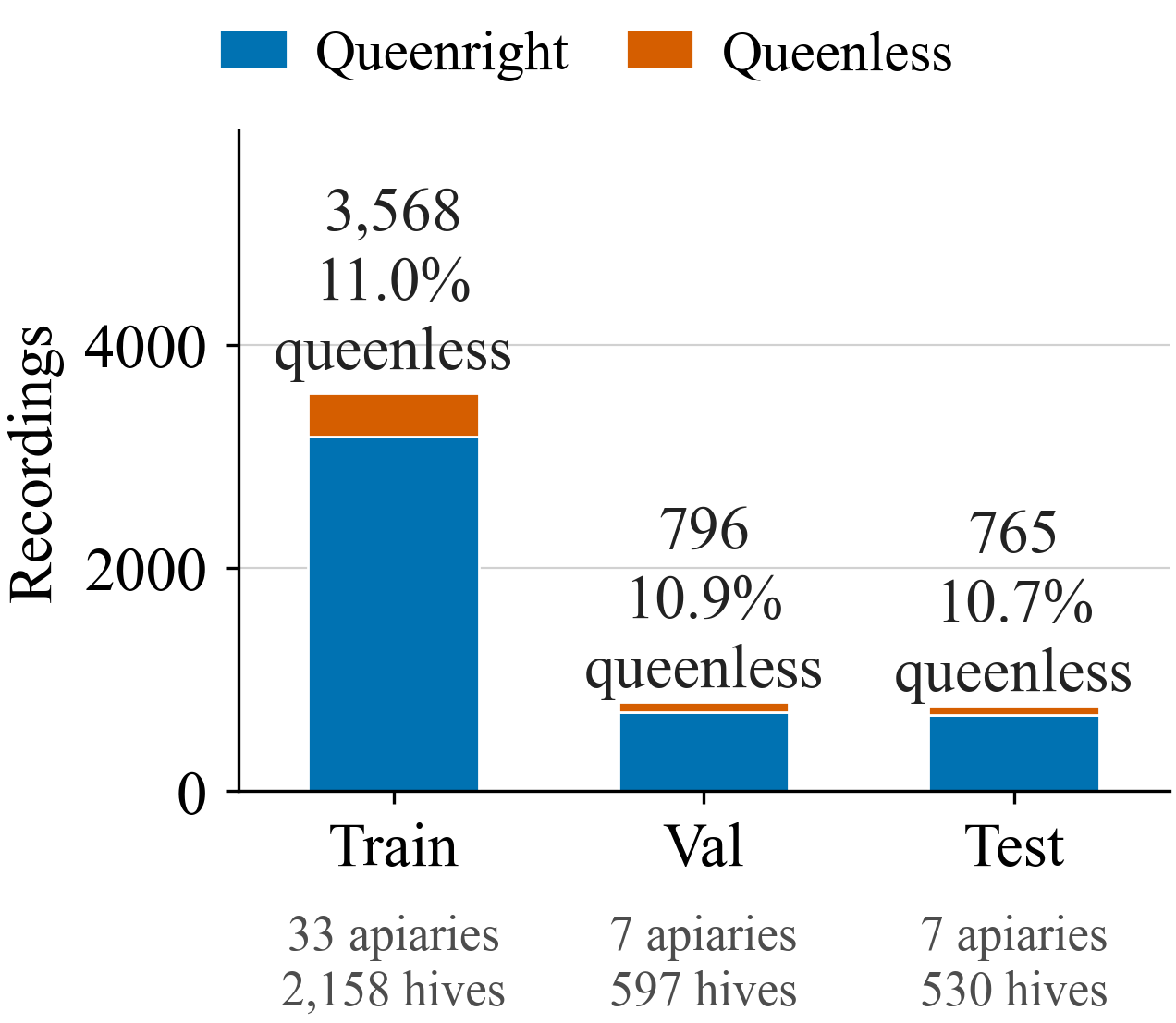}
}
\caption{Split composition under the two grouping keys: a) hive-independent and b) apiary-independent.}
\label{fig:split}
\end{figure}

\begin{table}
\centering
\caption{Training configuration. All representations of
Table~\ref{tab:features} share the CNN setting; the pretrained transformers use their own front-end and optimizer.}
\label{tab:cnn}
\small
\setlength{\tabcolsep}{5pt}
\begin{tabular}{@{}ll@{}}
\toprule
\multicolumn{2}{@{}l}{\textit{CNN (all representations)}} \\
\midrule
Backbone     & MobileNetV1, 10 blocks, layer norm \\
Pooling      & GeM ($p$ learned) $\rightarrow$ 512-d \\
Fusion       & $\oplus$ time 16-d, GPS 16-d $\rightarrow$ 544-d \\
Head         & dropout 0.6 $\rightarrow$ softmax (2) \\
Optimizer    & Adam, lr $10^{-3}$, batch 128, $\le$100 epochs \\
Weight decay & $L_2 = 5\times10^{-4}$ \\
Augmentation & noise $\sigma=0.05$, time roll $5\%$ \\
Schedule     & lr $\times0.7$ / 12 stagnant epochs, floor $10^{-6}$ \\
Stopping     & patience 10 on val.\ loss \\
\midrule
\multicolumn{2}{@{}l}{\textit{Pretrained transformers}} \\
\midrule
AST          & ViT-B, AudioSet-supervised; linear head \\
MSM-MAE      & ViT-B, self-supervised; LN\,$\rightarrow$\,256 GELU, drop.\ 0.3 \\
Input        & raw waveform, model's own log-mel front-end \\
Optimizer    & AdamW, linear decay, no weight decay \\
             & lr $10^{-5}$ \\
Batch        & 8 (AST), 16 (MSM-MAE); $\le$10 epochs \\
Stopping     & patience 3 on val.\ loss \\
\midrule
\multicolumn{2}{@{}l}{\textit{Shared}} \\
\midrule
Loss         & cross-entropy, class weights 4.90 / 0.56 \\
Selection    & best val.-loss epoch, restored for test \\
\bottomrule
\end{tabular}
\end{table}

\section{Results and Discussion}

Table~\ref{tab:main-results} compares the queenless detection performance under hive- and apiary-independent splits for the different representations and models. Overall, the modulation-based representations consistently outperform conventional spectral features. Under the hive-independent split, the 2D modulation spectrogram achieves the best performance, with an AUROC of 0.81 and AUPRC of 0.37, compared with 0.70 and 0.24 for MFCCs. It also substantially improves queenless recall from 0.31 to 0.67, indicating that temporal modulation patterns capture information relevant to queen status that is less effectively represented by conventional frame-level features. The 3D modulation representation provides no meaningful improvement over the simpler 2D representation.

Performance decreases considerably under the more challenging apiary-independent split, with the best modulation model reaching an AUROC of 0.63 and AUPRC of 0.26. Nevertheless, modulation representations remain the strongest features across both splits. The reduction in performance suggests substantial distribution shifts between apiaries, potentially caused by differences in hive environments, queen lineage, microphone placement, colony management, or background acoustic conditions. This highlights the importance of apiary-independent evaluation for assessing the practical generalization of acoustic monitoring systems.

Pre-trained encoders show mixed results. MSM-MAE performs competitively under the hive-independent split (AUROC = 0.79, AUPRC = 0.32), whereas AST achieves substantially lower performance and almost no queenless recall (0.01). MSM-MAE also degrades under the operation-independent split (AUROC = 0.57), remaining below the modulation-based CNNs. These results suggest that generic pre-trained representations do not necessarily provide an advantage for this task, while task-specific modulation representations better capture acoustic patterns associated with queen status. Overall, the results demonstrate the potential of modulation-based features for queenless detection while highlighting cross-operation generalization as a key remaining challenge.

\begin{table}
\centering
\caption{Queenless detection under two group-independent splits.}
\label{tab:main-results}
\footnotesize                      

\setlength{\tabcolsep}{3pt}
\renewcommand{\arraystretch}{0.95}

\begin{tabular}{l l c c c c c}
\toprule
\textbf{Feature} & \textbf{Model}
& \textbf{AUROC} & \textbf{AUPRC$_{\text{ql}}$} & \textbf{Prec.$_{\text{ql}}$} & \textbf{Rec.$_{\text{ql}}$} & \textbf{Rec.$_{\text{qr}}$} \\
\midrule
\multicolumn{7}{c}{\textbf{(a) Hive-independent split}} \\
\midrule
\multicolumn{7}{l}{\emph{Models trained from scratch on our data}} \\
\midrule
MFCC                        & \multirow{5}{*}{CNN} & 0.70 & 0.24 & 0.25 & 0.31 & 0.89 \\
Mel-spec             &                      & 0.75 & 0.26 & 0.25 & 0.30 & 0.89 \\
Multi-res Mel                 &                      & 0.76 & 0.33 & 0.44 & 0.31 & 0.95 \\
Mod-spec (2D) &                      & \textbf{0.81} & \textbf{0.37} & \textbf{0.30} & \textbf{0.67}  & \textbf{0.81} \\
Mod-spec (3D) &                      & 0.81   & 0.36 & 0.30 & 0.64   & 0.82 \\
\midrule
\multicolumn{7}{l}{\emph{Pre-trained encoders fine-tuned on our data (fixed log-mel input)}} \\
\midrule
\multirow{2}{*}{\shortstack[l]{log-mel\\(encoder-fixed)}}
                            & AST     & 0.67 & 0.18 & 0.33 & 0.01 & 1.0 \\
                            & MSM-MAE & 0.79 & 0.32 & 0.29 & 0.46 & 0.86 \\
\midrule
\multicolumn{7}{c}{\textbf{(b) Apiary-independent split}} \\
\midrule
\multicolumn{7}{l}{\emph{Models trained from scratch on our data}} \\
\midrule
MFCC                        & \multirow{5}{*}{CNN} & 0.56  & 0.15 & 0.16 & 0.34 & 0.79\\
Mel-spec             &                      & 0.59 & 0.18 & 0.16 & 0.52 & 0.66 \\
Multi-res Mel                 &                      & 0.58 & 0.18 & 0.25 & 0.26 & 0.91 \\
Mod-spec (2D) &                      & \textbf{0.63}   & \textbf{0.26} & \textbf{0.42} & \textbf{0.27}   & \textbf{0.96} \\
Mod-spec (3D)  &                      & 0.62   & 0.26 & 0.35 & 0.27   & 0.94 \\
\midrule
\multicolumn{7}{l}{\emph{Pre-trained encoders fine-tuned on our data (fixed log-mel input)}} \\
\midrule
\multirow{2}{*}{\shortstack[l]{log-mel\\(encoder-fixed)}}
                            & AST     & 0.52 & 0.16 &  0.10 &  0.55 & 0.41 \\
                            & MSM-MAE & 0.57 & 0.19 & 0.14 & 0.40 & 0.69 \\
\bottomrule
\end{tabular}
\end{table}

\section{Conclusions}

This study evaluated conventional, modulation-based, and pretrained audio representations for honeybee queen-state detection under hive- and apiary-independent splits. Modulation spectrograms consistently provided the strongest performance, particularly for unseen hives. However, performance declined under the more challenging apiary-independent setting, highlighting substantial cross-operation distribution shifts and the need for more robust generalization.



\bibliographystyle{IEEEbib}
\bibliography{refs}

\end{document}